\documentclass[aps,pra,showpacs,twocolumn,superscriptaddress]{revtex4-1}
\usepackage{graphicx}
\usepackage[usenames]{color}
\usepackage{amssymb,amsmath}
\newcommand{\eq}[1]{Eq.~(\ref{#1})}
\newcommand{\fig}[1]{Fig.~\ref{#1}}
\newcommand{\be}[1]{\begin{equation}\label{#1}}
\newcommand{\ee}{\end{equation}}

\usepackage[LGRgreek]{mathastext}

\begin{document}

\title{Controlling electron-electron correlation in frustrated double ionization of molecules with orthogonally polarized two-color laser fields }

%\author{A. Chen}
%\affiliation{Department of Physics and Astronomy, University College London, Gower Street, London WC1E 6BT, United Kingdom}
%\author{C. Lazarou}
%\affiliation{Department of Physics and Astronomy, University College London, Gower Street, London WC1E 6BT, United Kingdom}
%\author{H. Price}
%\affiliation{Department of Physics and Astronomy, University College London, Gower Street, London WC1E 6BT, United Kingdom}
%\author{A. Staudte}
%\affiliation{Joint Laboratory for Attosecond Science, University of Ottawa and National Research Council,
%100 Sussex Drive, Ottawa, Ontario, Canada K1A 0R6}
%\author{I. Ben-Itzhak}
%\affiliation{Kansas State University, Manhattan}

\author{A. Chen}
\affiliation{Department of Physics and Astronomy, University College London, Gower Street, London WC1E 6BT, United Kingdom}
\author{M. F. Kling}
\affiliation{Department of Physics, Ludwig-Maximilians-Universit\"{a}t Munich, Am Coulombwall 1, D-85748 Garching, Germany}
\affiliation{Max Planck Institute of Quantum Optics, Hans-Kopfermann-Str. 1, D-85748 Garching, Germany}
\author{A. Emmanouilidou}
\affiliation{Department of Physics and Astronomy, University College London, Gower Street, London WC1E 6BT, United Kingdom}

\begin{abstract}
We demonstrate the control of electron-electron correlation in frustrated double ionization (FDI) of the two-electron triatomic molecule D$_{3}^{+}$ when driven by two orthogonally polarized two-color laser fields. We employ a three-dimensional semi-classical model that fully accounts for the electron and nuclear motion in strong fields. We analyze  the FDI probability and the distribution of the momentum of the escaping electron along the polarization direction of the longer wavelength and more intense laser field. These observables when considered in conjunction bear clear signatures of the prevalence or absence of electron-electron correlation in FDI, depending on the time-delay between the two laser pulses. We find that D$_{3}^{+}$ is a better candidate compared to H$_{2}$ for demonstrating  also experimentally that electron-electron correlation indeed underlies FDI.
 \end{abstract}
\date{\today}

\maketitle

Frustrated double ionization (FDI) is a major process in the nonlinear response  of multi-center molecules when driven by intense laser fields, accounting roughly for 10\% of all ionization events \cite{Eichmann1,Emmanouilidou2012}. 
In frustrated ionization an electron first tunnel-ionizes in the driving laser field.  Then, due to the electric field of the laser pulse, it is recaptured by the parent ion in a Rydberg state \cite{Eichmann2}. This process is a candidate for the inversion of N$_{2}$ in  free-space air lasing \cite{lasing}.
In FDI an electron escapes and another one  occupies a Rydberg state at the end of the laser pulse.   FDI has attracted considerable interest in recent years in a number of experimental studies  in the context of $\mathrm{H_{2}}$ \cite{Eichmann1} and of the triatomic molecules $\mathrm{D_{3}^{+}}$ and $\mathrm{H_{3}^{+}}$ \cite{JcKenna,JcKenna2,JMcKenna}.
 
 In theoretical studies of strongly-driven two-electron diatomic and triatomic molecules, two pathways of FDI have been identified \cite{Emmanouilidou2012,Emmanouilidou2016}. Electron-electron correlation is important, primarily,  for one of the two pathways.
It is well accepted that electron-electron correlation underlies a significant part of  double ionization in strongly-driven molecules---a mechanism known as non-sequential double ionization  \cite{NSDI1,NSDI2}.
However,  electron-electron correlation in FDI has yet to be accessed experimentally. 

Here, we propose a road for future experiments to identify the important role of electron-electron correlation in FDI. 
We   identify the parameters of orthogonally polarized two-color (OTC) laser fields  that best control  the relevant pathway for electron-electron correlation in FDI. We demonstrate traces of attosecond control of electron motion in space and time  in two observables of  FDI as a function of the time-delay between the fundamental 800 nm and the second harmonic 400 nm laser field. We show that, together, the FDI probability   and the momentum of the escaping electron along the fundamental  laser field bear clear signatures of the turning on and off of electron-electron correlation. 

Two-color laser fields  are an efficient tool for controlling electron motion \cite{Kitzler,Richter} and for steering the outcome of chemical reactions  \cite{Ray,Li,Wu}. Other applications include  the field-free orientation of molecules \cite{De,Frumker,Znakovskaya}, the generation of high-harmonic spectra \cite{HHGOTC3,HHGOTC1,Brugnera,HHGOTC2} and  probing atomic and molecular orbital symmetry \cite{Dudovic,Shafir, Niikura2}.

The strongly-driven dynamics of two electrons and three nuclei poses a challenge for fully ab-initio quantum mechanical calculations. The latter techniques can currently address one electron triatomic molecules \cite{Bandrauk}. Therefore, we rely on classical and semi-classical models for understanding the fragmentation dynamics in triatomic molecules driven by intense infrared laser pulses \cite{classic1, Emmanouilidou2016}. Our work employs a three-dimensional  semi-classical model. This model has provided significant insights into  FDI  for  strongly-driven $\mathrm{H_{2}}$ \cite{Emmanouilidou2012} and $\mathrm{D_{3}^{+}}$ \cite{Emmanouilidou2016}. Our previous result for the distribution of the kinetic energy release of the Coulomb exploding  nuclei  in FDI of D$_{3}^{+}$ was  in good agreement with experiment \cite{JMcKenna}.

We employ the initial state of $\mathrm{D_{3}^{+}}$  that is accessed experimentally via the reaction $\mathrm{D_{2}+D_{2}^{+}\rightarrow D_{3}^{+}+D}$ \cite{JMcKenna,JcKenna}.
It consists of a superposition of triangular-configuration vibrational states ${\nu=1-12}$ \cite{D3_vibrational, JMcKenna}.   We assume that most of the D$_3^+$ ionization occurs at the outer classical turning point of the vibrational levels \cite{Ergler2006PRL,Goll2006PRL}. The turning point varies from 2.04~a.u. ($v=1$) to 2.92~a.u. ($v=12$)  \cite{D3_vibrational, D3_potential}. We initialize the nuclei at rest for all vibrational levels, since an initial pre-dissociation does not significantly modify the ionization dynamics \cite{Emmanouilidou2014}.

The combined strength of the two laser fields is  within the below-the-barrier ionization regime. To formulate the initial state of the two electrons, we assume that one electron (electron 1) tunnel-ionizes at time $\mathrm{t_{0}}$  in the field-lowered Coulomb potential. For this quantum-mechanical step, we compute the ionization rate using a semi-classical formula \cite{Murray2011}. t$_{0}$ is selected using importance sampling \cite{importance_samp} in the time interval the two-color laser field is present. The ionization rate is then used as the importance sampling distribution. For electron 1, the velocity component that is transverse to the OTC laser fields is given by a Gaussian \cite{ADK} and the component that is parallel is set equal to zero. The initial state of the initially bound electron (electron 2) is described by a microcanonical distribution \cite{microcanonical}.

Another quantum mechanical aspect of our 3D model is tunneling of each electron during the propagation with a probability given by the Wentzel-Kramers-Brillouin approximation    \cite{Emmanouilidou2012, Emmanouilidou2014}. This aspect is essential to accurately describe the enhanced ionization process  \cite{NSDI2,bandrauk1996}. In EI, at a critical distance of the nuclei, a double potential well is formed such that it is easier for an electron bound to the higher potential well to tunnel to the lower potential well and subsequently ionize. The time propagation is classical, starting from time $\mathrm{t_{0}}$. We solve the classical equations of motion for the Hamiltonian of the strongly-driven five-body system, while fully accounting for the Coulomb singularities  \cite{Emmanouilidou2014}.

 The OTC laser field we employ   is of the form
\begin{eqnarray}
{\bf E}(t,\Delta t)&=&E_{\omega}f(t)cos(\omega t)\hat{z}+E_{2\omega}f(t+\Delta t)cos[2\omega (t+\Delta t)]\hat{x}\nonumber\\
&&f(t)=exp\left(-2ln2\left(\frac{t}{\tau_{FWHM}}\right)^2\right),
\label{eq1}
\end{eqnarray}
with $\omega=0.057$ a.u. for commonly used Ti:sapphire lasers at 800\,nm. $T_{\omega}$ and $T_{2\omega}$ are the corresponding periods of the fundamental  and second harmonic laser fields, polarized along the $\hat{z}$- and $\hat{x}$-axis, respectively. $\tau_{FWHM}=40$ fs is the full-width-half-maximum. $\Delta t$ is the time delay between the $\omega-2\omega$ pulses.
We consider $E_{\omega}=0.08$ a.u., since for this field strength pathway B of FDI,  where electron-electron correlation is present, prevails over pathway A---4.8\% versus 3.6\%  \cite{Emmanouilidou2016}.

In FDI of $\mathrm{D_{3}^{+}}$ the final fragments are a neutral excited fragment $\mathrm{D^{*}}$, two $\mathrm{D^{+}}$ ions and one escaping electron. In the neutral excited fragment $\mathrm{D^{*}}$ the electron occupies a Rydberg state with quantum number $\mathrm{n>1}$. The difference between the two FDI pathways lies in how fast the ionizing electron escapes following the turn on of the laser field \cite{Emmanouilidou2012}. In pathway A, electron 1 tunnel-ionizes and  escapes early on. Electron 2 gains energy from  an EI-like process and tunnel-ionizes. It does not have enough drift energy to escape when the laser field is turned off and finally it occupies a Rydberg state, $\mathrm{D^{*}}$. In pathway B, electron 1 tunnel-ionizes and quivers in the laser field returning to the core. Electron 2 gains energy from both  an EI-like process and the returning electron 1 and tunnel-ionizes after a few periods of the laser field. When the laser field is turned off,  electron 1 does not have enough energy to escape and remains bound in a Rydberg state. It follows that electron-electron correlation is more pronounced in pathway B \cite{Emmanouilidou2012}.

To compute the FDI probability as a function of the time delay $\Delta t$ of the $\omega-2\omega$ pulses, we use
\begin{eqnarray}
P^{FDI}(\Delta t)=\frac{\sum_{\nu,i} P_{\nu} \Gamma(\Delta t,\nu,i)P^{FDI}(\Delta t,\nu,i)}{\sum_{\nu,i}P_{\nu} \Gamma(\Delta t,\nu,i)},
\label{prob}
\end{eqnarray}
where $i$ refers to the different orientations of the molecule with respect to the $z$-component of the laser field. We consider only two cases of planar alignment, that is, one side of the equilateral, molecular triangle is either parallel or perpendicular to the $\mathrm{\hat{z}-}axis$. $\Gamma(\Delta t,\nu,i)$ is given by
\begin{eqnarray}
\Gamma(\Delta t, \nu, i)=\int_{t_{i}}^{t_{f}} \Gamma(t_{0}, \Delta t,\nu,i) dt_{0},
\end{eqnarray}
where the integration is over the duration of the OTC field. $\Gamma(t_{0}, \Delta t,\nu,i)$ is the ionization rate at time $\mathrm{t_{0}}$ for a certain molecular orientation $i$, vibrational state $\nu$ and time delay $\Delta t$.
P$_{\nu}$ is the percentage of the vibrational state $\nu$ in the initial state of $\mathrm{D_{3}^{+}}$ \cite{D3_vibrational}. P$^{FDI}(\Delta t,\nu,i)$ is the number of FDI events out of all initiated classical trajectories for a certain molecular orientation $i$, vibrational state $\nu$ and time delay $\Delta t$. Due to the challenging computations involved, we approximate \eq{prob} using the $\nu=8$ state of $\mathrm{D_{3}^{+}}$. This approximation is justified, since we find that the $\nu=8$ state contributes the most in the sum in \eq{prob}. We obtain very similar results for the  $\nu=7,9$ states,  which contribute to the sum in \eq{prob} less than  the $\nu=8$  state but more than the other states.

% For $\mathrm{E_x=}$0.02 a.u., we find  that the FDI probability has a small variation  with $\Delta t$ (not shown). This small change is  due to pathway B of FDI, with the probability of pathway B  varying around 4.9\% as a function of $\Delta t$.  The  probability of pathway A  remains roughly constant and equal to 3.9\%.  3.9\% and 4.9\% are the probabilities  of pathway A and B of FDI for  $\mathrm{E_x=0}$ a.u..
   \begin{figure}
      \begin{center}
  \centering
  \includegraphics[clip,width=0.50\textwidth]{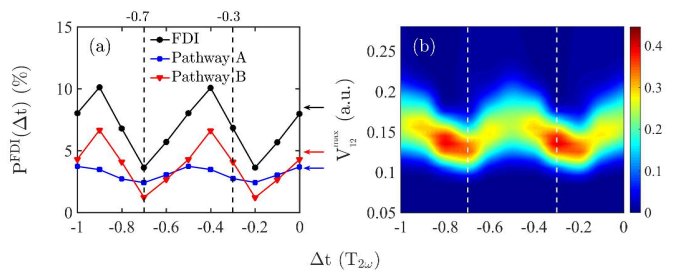}
   \caption[]{
  (a) The FDI probability and the  probabilities of pathways A and B and (b) the distribution of $\mathrm{V_{12}^{max}}$ are plotted as a function of  $\Delta t$ for $E_\omega=0.08$ a.u. and $E_{2\omega}=0.05$ a.u.. In (a) the arrows on the right indicate the corresponding probabilities when $E_{2\omega}=0$ a.u.. }
   \label{figure1}
  \end{center}
  \end{figure}
In \fig{figure1}(a), for E$_{2\omega}=0.05$ a.u., we plot the FDI probability as a function of the time delay for $\Delta t \in [0,T_{2\omega}]$. The results are periodic with $T_{2\omega}/2$. We find that the FDI probability changes significantly with $\Delta t$.   This change is mainly due to pathway B with probability  that varies from 1.2\% at $\Delta t=$-0.2, -0.7 T$_{2\omega}$ to 6.7\% at $\Delta t=$-0.4, -0.9 T$_{2\omega}$. In contrast, the probability of pathway A changes significantly less varying from  2.4\% to 3.7\%. For E$_{2\omega}<$0.05 a.u., the probability of  pathway B varies less than for E$_{2\omega}=0.05$ a.u.. 

 Control of electron-electron correlation in double ionization in atoms has been demonstrated through the free parameters  $\Delta t$ and E$_{2\omega}$ of OTC laser fields  \cite{Zhou10,NSDIOTC2,NSDIOTC,DIOTC,Mancuso16,Eckart16}. The time-delay between the laser fields can significantly affect the time and the distance of the closest approach of the returning electron \cite{Kitzler}. For FDI,  this is demonstrated in \fig{figure1}(b). For each classical trajectory labelled as FDI, we compute the  maximum of the Coulomb potential energy $\mathrm{1/|{\bf r}_{1}-{\bf r}_2|}$, $\mathrm{V_{12}^{max}}$. Then, we plot the distribution of $\mathrm{V_{12}^{max}}$ as a function of $\Delta t$. The minimum values of $\mathrm{V_{12}^{max}}$ correspond to electron 1 being at a maximum distance from the core, i.e.  minimum electron-electron correlation. Comparing \fig{figure1}(a) with (b), we find that these minima occur at the same $\Delta t$s, where the FDI probability and the probability of pathway B  is minimum, i.e. at $\Delta t=$-0.2, -0.7 T$_{2\omega}$.

The probability of each FDI pathway as well as $\mathrm{V_{12}^{max}}$ are not experimentally accessible quantities. To demonstrate the presence of electron-electron correlation in FDI, in addition to the sharp change of the FDI probability with $\Delta t$, we need one more experimentally accessible observable. This observable should bare clear signatures of the prevalence of pathway A at the $\Delta t$s where the minima of the FDI probability occur, i.e. at $\Delta t=$-0.2, -0.7 T$_{2\omega}$.
We find that such an FDI-observable is the change of the momentum of the escaping electron along the polarization direction of the fundamental ($\omega$) laser field, $\mathrm{p_{z}}$, with $\Delta t$.
  \begin{figure} [h]
  \begin{center}
  \centering
  \includegraphics[clip,width=0.530\textwidth]{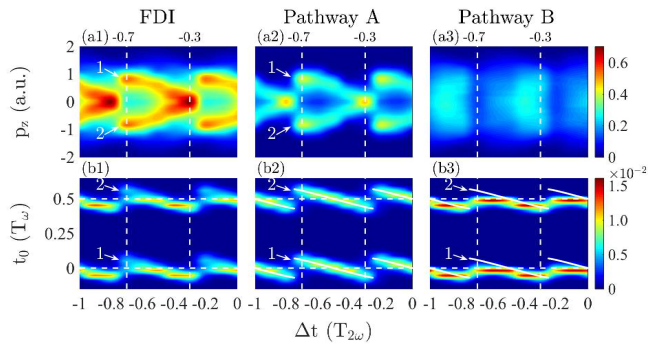}
  \caption[]{The distribution of p$_{z}$ for FDI (a1) and for pathways A (a2) and B (a3) are plotted as a function of  $\Delta t$. For each $\Delta t$, the distribution of p$_z$ for FDI is normalized to 1 while for pathways A and B  it is normalized with respect to the total FDI probability. The distribution of the time electron 1 tunnel-ionizes during half cycles 1 and 2  for FDI (b1) and for pathways A (b2) and B (b3) is plotted as a function of  $\Delta t$. For each $\Delta t$, the distribution of t$_{0}$  in (b1)-(b3) is normalized to 1.   t$_{max}$ is plotted with white dots (appear as white lines) in (b2) and (b3).}
  \label{figure3}
  \end{center}
  \end{figure}

\begin{figure}
  \begin{center}
  \centering
  \includegraphics[clip,width=0.48\textwidth]{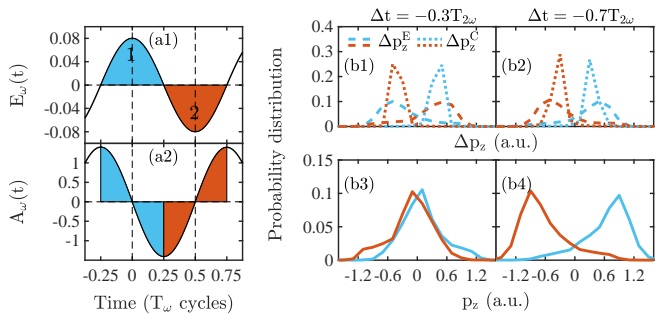}
  \caption[]{  Half cycles 1 and 2  for E$_{\omega}$ (a1) and its vector potential (a2). For pathway A, the distributions of $\Delta p_{z}^{E}$ and $\Delta p_{z}^{C}$ are plotted for half cycles 1 and 2  for  $\Delta t=$-0.3 T$_{2\omega}$ (b1) and  $\Delta t=$-0.7 T$_{2\omega}$ (b2).
  The distribution of p$_{z}$ is plotted for half cycles 1 and 2 for  $\Delta t=$-0.3 T$_{2\omega}$ (b3) and  $\Delta t=$-0.7 T$_{2\omega}$ (b4).}
  \label{figure4}
  \end{center}
\end{figure}

In \fig{figure3}(a1) we plot the distribution of p$_{z}$ as a function of $\Delta t$ for one period of the results, that is, in the interval $\Delta t\in[-0.7T_{2\omega}, -0.2T_{2\omega}]$ in steps of $\Delta t=0.1$ T$_{2\omega}$. We find that the distribution of p$_{z}$ has a V-shape. It consists of two branches that have a maximum split at $\Delta t=$-0.7 T$_{2\omega}$, with peak values of p$_{z}$ around -0.85 a.u. and 0.85 a.u.. The two branches coalesce at $\Delta t=$-0.3 T$_{2\omega}$, with p$_z$ centered around zero. Moreover, FDI events with electron 1 tunnel-ionizing during half cycles with extrema at nT$_\omega$ (n/2T$_{\omega}$) contribute to the upper (lower) branch of the distribution of p$_z$. n takes both positive and negative integer values. We find that half cycles 1 and 2, see \fig{figure4}(a1) and (a2), with extrema at 0  and T/2 of the E$_{\omega}$ laser field, respectively, contribute the most to the momentum distribution of p$_{z}$. Thus, it suffices to focus our studies on half cycles 1 and 2.

First, we investigate the change of the distribution of the time electron 1 tunnel-ionizes t$_{0}$ with $\Delta t$, see  \fig{figure3}(b1). When the second harmonic ($2\omega$) field is turned off, t$_{0}$ is centered around the extrema of half cycles 1 and 2 (not shown). However, when the $2\omega$-field is turned on, depending on $\Delta t$, electron 1 tunnel-ionizes at times t$_{0}$ that are shifted to the right or to the left of the extrema of half cycles 1 and 2, see \fig{figure3}(b1). Moreover, we find that t$_{0}$ shifts monotonically from the lowest value of the shift at $\Delta t=-0.3$ T$_{2\omega}$ to its highest value at $\Delta t=-0.7$ T$_{2\omega}$. We find that this change of t$_{0}$ is due to the monotonic change with $\Delta t$ of the time t$_{max}$ when the  magnitude of the OTC laser field is maximum. That is, for each $\Delta t$, we compute the time t$_{max}$ when the laser field in \eq{eq1} is maximum. t$_{max}$ is also the time that the ionization rate is maximum. We plot t$_{max}$ for half cycles 1 and 2 in \fig{figure3}(b2) and (b3). We compare t$_{max}$ with the distribution of t$_{0}$ for pathways A and B.  We find t$_{max}$ to be closest to the distribution of t$_{0}$ for pathway A. Indeed, only when electron 1 is the escaping electron will the time electron 1 tunnel-ionizes  be roughly equal to the time the ionization rate is maximum. In pathway B it is electron 2 that escapes. Thus, the time t$_{0}$  must be such that  both the ionization rate and the electron-electron correlation efficiently combine  to ionize electron 2.

Next, for pathway A, we explain how the two brunches of the distribution of p$_{z}$ split  when t$_{0}$ shifts to the right  of the extrema of  half cycles 1 and 2 ($\Delta t$=-0.7 $T{_{2\omega}}$) or coalesce
 when t$_{0}$ shifts to the left  ($\Delta t=$-0.3 T$_{2\omega}$).
      %different  contribution to p$_{z}$ of each of the half cycles 1 and 2  and thus a splitting of the momentum distribution at $\Delta t=$-0.7 T$_{x}$ while shift of t$_{0}$ to the left of the extremum of half cycles 1 and 2  results in a similar  contribution to p$_{z}$ of each of the half cycles 1 and 2  and thus a coalescence  of the momentum distribution at $\Delta t=$-0.3 T$_{x}$.
We compute the changes in p$_{z}$ of the escaping electron 1 due to the $\omega$-field as well as due to the interaction of electron 1 with the core. These  momentum changes are given by
{\small
\begin{eqnarray}
\Delta p_{z}^{E}(\Delta t, t_{0})&=&\int_{t_{0}}^{\infty}-E_{\omega}(t)dt, \nonumber\\
\Delta p_{z}^{C}(\Delta t,t_{0})&=&\int_{t_{0}}^{\infty}\left(\sum_{i=1}^{3}\frac{{\bf R_{i}}-{\bf r}_{1}}{|{\bf r}_{1}-{\bf R}_{i}|^{3}}+\frac{{\bf r}_{1}-{\bf r}_{2}}{|{\bf r}_{1}-{\bf r}_{2}|^3}\right)\cdot{\hat z}dt,
\end{eqnarray}
}with ${\bf R}_{i}$ the position of the nuclei.  Using the times t$_{0}$ for the events labeled as pathway A, we plot the probability distributions of $\Delta p_{z}^{E}$ and of $\Delta p_{z}^{C}$ at $\Delta t=$-0.3 T$_{2\omega}$ and at $\Delta t=$-0.7 T$_{2\omega}$ in \fig{figure4}(b1) and (b2), respectively. We find that, for both $\Delta t$s, the distribution of $\Delta p_{z}^{C}$ peaks at positive  (negative) values of $\Delta p_{z}^{C}$ when electron 1 tunnel-ionizes during half cycle 1 (2). Indeed, during half cycle 1 (2), electron 1 tunnel-ionizes to the left (right) of the field-lowered Coulomb potential. Then, the force from the core acts along  the positive (negative) $\hat{z}$-axis resulting in the distribution $\Delta p_{z}^{C}$ peaking around positive (negative) values for half cycle 1 (2). We find that the contribution of the electron-electron repulsion term  is small compared to the attraction from the nucleus in $\Delta p_{z}^{C}$. In contrast, the distribution of $\Delta p_{z}^{E}$ peaking at positive or negative values of  $\Delta p_{z}^{E}$  depends on whether $t_{0}$ shifts to the right or to the left of the extrema of  half cycles 1 and 2, i.e. it depends on $\Delta t$. For $\Delta t$=-0.3 T$_{2\omega}$, when t$_{0}$ shifts to the left of the extrema of half cycles 1 (2), the vector potential is positive (negative) resulting in the distribution of $\Delta p_{z}^{E}$ peaking at negative (positive) values of $\Delta p_{z}^{E}$. Similarly, for $\Delta t$=-0.7 T$_{2\omega}$, the distribution of $\Delta p_{z}^{E}$ peaks at positive (negative) values of  $\Delta p_{z}^{E}$ for half cycle 1 (2).

 In \fig{figure3}(b3) and (b4), we plot the distributions of the final momentum p$_{z}$, which is given by $\Delta p_{z}^{E}$+$\Delta p_{z}^{C}$+p$_{z,t_{0}}$.
 The distribution of  the component of the initial momentum of electron 1, p$_{z,t_{0}}$,  has a small contribution to p$_{z}$ and is not shown. In \fig{figure4}(b3), for $\Delta t=$-0.3 T$_{2\omega}$, we show that the distributions of p$_{z}$ for half cycles 1 and 2 are similar and peak at zero. They give rise  to the two branches of the distribution p$_{z}$ coalescing in \fig{figure3}(a2) and (a1). In \fig{figure4}(b4), for $\Delta t=$-0.7 T$_{2\omega}$, we find that the distributions of p$_{z}$ for half cycles 1 and 2 are quite different with peaks at 0.85 a.u. and -0.85 a.u., respectively. They  give rise to the split of the two branches of the distribution p$_{z}$ in \fig{figure3}(a2) and (a1). Unlike pathway A, for pathway B the distribution of p$_{z}$ as a function of $\Delta t$ in \fig{figure3}(a3) is very broad. The reason is that electron 2 has time to interact with the core since it tunnel-ionizes after a few cycles of the laser field.

\begin{figure}
  \begin{center}
  \centering
  \includegraphics[clip,width=0.480\textwidth]{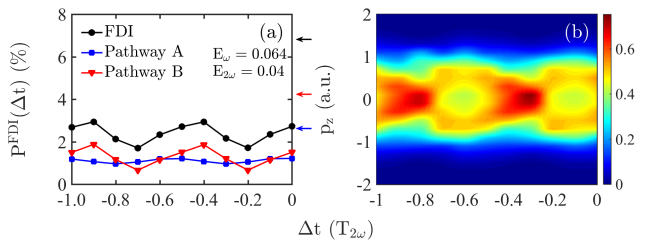}
  \caption[]{(a) and (b) similar to  \fig{figure1}(a) and  \fig{figure3}(a1), respectively,  for H$_{2}$.}
  \label{figure_H2}
  \end{center}
\end{figure}

Finally, we show that a similar level of control of electron-electron correlation with OTC fields can not be achieved for  H$_{2}$. We choose E$_{\omega}=0.064$ a.u. so that E$_{\omega}$ for H$_{2}$ and D$_{3}^{+}$  has the same percentage difference from the field strength that corresponds to over-the-barrier ionization. We choose E$_{2\omega}=$0.04 a.u. so that E$_\omega$/E$_{2\omega}$ is the same  for both molecules. We show in \fig{figure_H2}(a) that, for all $\Delta t$s, the FDI probability significantly reduces when the $2\omega$-field is turned on. Indeed, its maximum value is 2.7\% compared to 6.8\% for E$_{2\omega}=0$ a.u.. In contrast, in D$_{3}^{+}$  the FDI probability changes from  8.5\% without $2\omega$-field to a maximum value of 10.5\% for E$_{2\omega}=0.05$ a.u.. We find that the FDI probability as well as the probability of pathway B do not significantly change with $\Delta t$. In addition, the two branches of the  V-shaped distribution p$_{z}$ of the escaping electron are not as pronounced in \fig{figure_H2}(b) as for D$_{3}^{+}$. The results in \fig{figure_H2} are obtained when the inter-nuclear axis of H$_{2}$ is parallel to E$_{\omega}$.  We find similar results for  a perpendicular orientation, however, for E$_{2\omega}=0$ a.u.,  the FDI probability is almost zero.

In conclusion, we have shown that control of electron-electron correlation in FDI can be achieved employing OTC fields in D$_{3}^{+}$. We find that the FDI probability changes sharply with the time-delay between the two laser fields. Moreover, we identify a split in the distribution of the final momentum of the escaping electron that  takes place at  time-delays where the FDI probability is minimum. We show this split to be a signature of the absence of electron-electron correlation. It then follows that electron-electron correlation is present for the time-delays, where the FDI probability is maximum. Future experiments can employ our scheme to demonstrate the importance of electron-electron correlation in FDI.

%\begin{acknowledgments}
A. E. acknowledges  the EPSRC grant no. J0171831 and the use of the computational resources of Legion at UCL. M.F.K acknowledges support by the German Research Foundation (DFG) via the Cluster of Excellence: Munich Centre for Advanced Photonics (MAP), and by the European Union (EU) via the European Research Council (ERC) grant ATTOCO.
%\end{acknowledgments}

\begin{flushleft}
\small
%\section*{References}
%\small
%\bibliographystyle{aipauth4-1}
%\bibliographystyle{unsrt}
%\bibliographystyle{apalike}
%\bibliographystyle{iopart-num}
\end{flushleft}

%\bibliography{cite}

\end{document}